\documentclass[a4paper,11pt]{article}
\pdfoutput=1

\usepackage{jheppub}

\usepackage{graphicx}
\usepackage{amsmath}
\usepackage{amssymb}
\usepackage{amsfonts}
\usepackage{dcolumn}
\usepackage{bm}
\usepackage[dvipsnames]{xcolor}
\usepackage[utf8]{inputenc}
\usepackage{subfigure}
\usepackage{gensymb}
\usepackage{cleveref}
\usepackage{colortbl}
\usepackage{tabularx}
\usepackage{tabulary}

\usepackage[T1]{fontenc}
\usepackage{pstricks}
\usepackage{color}
\usepackage{multirow}
\usepackage{slashed}
\usepackage{mathtools}
\usepackage{mathrsfs}
\usepackage{bbold}
\usepackage[force]{feynmp-auto}
\usepackage{verbatim}

\newcommand{\Hub}{\mathcal{H}}

\DeclareMathAlphabet{\mathpzc}{OT1}{pzc}{m}{it}

\hypersetup{colorlinks,linkcolor={blue},citecolor={teal},urlcolor={violet}}

\newcommand{\SOTON}{Department of Physics and Astronomy, University of Southampton, SO17 1BJ Southampton, United Kingdom}

\begin{document}

\title{Cosmic string gravitational waves from global $U(1)_{B-L}$ symmetry breaking as a probe of the type I seesaw scale}

\author[a,1]{Bowen Fu,\note{\url{https://orcid.org/0000-0003-2270-8352}}}
\author[b,2]{Anish Ghoshal\note{\url{https://orcid.org/0000-0001-7045-302X}}}
\author[a,3]{Stephen F. King\note{\url{https://orcid.org/0000-0002-4351-7507}}}

\affiliation[a]{\SOTON}
\affiliation[b]{Institute of Theoretical Physics, Faculty of Physics, University of Warsaw, ul. Pasteura 5, 02-093 Warsaw, Poland}

\emailAdd{B.Fu@soton.ac.uk}
\emailAdd{anish.ghoshal@fuw.edu.pl}
\emailAdd{king@soton.ac.uk}

\date{\today}

\abstract{
In type I seesaw models, the right-handed neutrinos are typically super-heavy, consistent with the generation of baryon asymmetry via standard leptogenesis. Primordial gravitational waves of cosmological origin provides a new window to probe such high scale physics, which would otherwise be inaccessible. By considering a {\em global} $U(1)_{B-L}$ extension of the type I seesaw model, we explore the connection between the heaviest right-handed neutrino mass and primordial gravitational waves arising from the dynamics of global cosmic string network. As a concrete example, we study a global $U(1)_{B-L}$ extension of the Littlest Seesaw model, and show that the inevitable GW signals, if detectable, probe the parameter space that can accommodate neutrino oscillation data and successful leptogenesis, while respecting theoretical constraints like perturbativity of the theory. Including CMB constraints from polarization and dark radiation leaves a large region of parameter space of the model, including the best fit regions, which can be probed by GW detectors like LISA and ET in the near future.
In general, the GW detectors can test high scale type I seesaw models with the heaviest right-handed neutrino mass above $2.5 \times 10^{14}$ GeV, assuming the perturbativity, and $7 \times 10^{13}$ GeV assuming that the coupling between the heaviest right-handed neutrino and the $U(1)_{B-L}$ breaking scalar is less than unity. 

%\ag{Should we add what new ingredients we studied in this paper just for the model side that were not done in any paper before other than cosmology ?}
}

\maketitle
\tableofcontents

%%%%%%%%%%%%%%%%%%%%%%%%%%%%%%%%%%%%%%%%%%%
\section{The introduction}
%%%%%%%%%%%%%%%%%%%%%%%%%%%%%%%%%%%%%%%%%%%

Evidenced by the neutrino oscillation experiments \cite{2016NuPhB.908....1O}, the existence of neutrino masses and mixing represents the most convincing physics beyond the Standard Model.  In the past half-century, theorists have invented hundreds of models to interpret the existence of the neutrino masses and most of them lead to an effective dimension-five Weinberg operator \cite{Weinberg:1979sa}. 
Among those models, the most popular and well-studied ones are the tree-level realisations of the Weinberg operator, namely the type I \cite{Minkowski:1977sc,Yanagida:1979as,GellMann:1980vs,Mohapatra:1979ia}, II \cite{Magg:1980ut,Schechter:1980gr,Wetterich:1981bx,Lazarides:1980nt,Mohapatra:1980yp,Ma:1998dx} and III \cite{Foot:1988aq,Ma:1998dn,Ma:2002pf,Hambye:2003rt} seesaw models. However, the most general version of seesaw model has many free parameters. In general, there are not enough physical constraints to fix the parameters, and thus the model is hard to be tested even indirectly. A natural and effective solution to reduce the number of free parameters is to consider only two right-handed neutrinos (2RHN) with one texture zero \cite{King:1999mb, King:2002nf}, in which the lightest neutrino has zero mass. The number of free parameters could be further reduced by imposing two texture zeros in the Dirac neutrino mass matrix \cite{Frampton:2002qc}. However, such a two texture zero model is incompatible with the normal hierarchy of neutrino masses even though the consistency with cosmological leptogenesis is kept \cite{Fukugita:1986hr,Guo:2003cc, Ibarra:2003up, Mei:2003gn, Guo:2006qa, Antusch:2011nz,Harigaya:2012bw, Zhang:2015tea}, while the one texture zero model is compatible with the normal neutrino mass hierarchy.

In neutrino mass models with 2RHN admitting some flavour structures, the model parameter can be very constrained, leading to a strong prediction on the seesaw scale. An example is the Littlest Seesaw (LS) model, which is based on the one texture zero 2RHN model with a constrained sequential dominance (CSD) form of Dirac neutrino mass matrix of level $n$ where $n=3$ leads to an excellent fit to low energy neutrino data \cite{King:2013iva, Bjorkeroth:2014vha, King:2015dvf,Bjorkeroth:2015ora,Bjorkeroth:2015tsa,King:2016yvg,Ballett:2016yod}. The number of independent Yukawa couplings is only two, which means the model is highly predictive. Following a fitting result for three degrees of freedom with the low-energy neutrino data and leptogenesis in such a model \cite{King:2018fqh}, a further extension has been made to the model to explain the existence of dark matter \cite{Chianese:2019epo,Chianese:2020khl}. However, in order to explain the baryon asymmetry through standard thermal leptogenesis, such kinds of models require the RHNs to be superheavy. The typical scale of the lightest RHN is around $10^{10}$ GeV, which is far beyond the reach of current or foreseen experiments. 

The recent discovery of gravitational wave events (astrophysical sources by the LIGO-Virgo collaboration in 2015~\cite{Abbott:2017xzu}) provides a new pathway to physics beyond the standard model (BSM), particularly as a window into the pre-BBN universe with several upcoming GW detectors expected in the near future such as LISA~\cite{Audley:2017drz}, BBO-DECIGO~\cite{Yagi:2011wg}, the 
Einstein Telescope~(ET)~\cite{Punturo:2010zz,Hild:2010id},
and Cosmic Explorer~(CE)~\cite{Evans:2016mbw}. Several studies in the recent years explored various interesting connection between BSM physics (involving neutrinos and leptogenesis) and gravitational waves of cosmological origin such as that from local cosmic strings \cite{Dror:2019syi}, domain walls \cite{Barman:2022yos} and other topological defects \cite{Dunsky:2021tih} or from nucleating and colliding vacuum bubbles \cite{Dasgupta:2022isg,Borah:2022cdx,Fu:2022eun,Borah:2023saq}, graviton bremmstrahlung \cite{Ghoshal:2022kqp} and primordial black holes \cite{Bhaumik:2022pil,Bhaumik:2022zdd,Borah:2022vsu}.
These previous studies on GW \cite{Vachaspati:1984gt,Buchmuller:2013lra,Chao:2017ilw,Okada:2018xdh,Buchmuller:2019gfy,Hasegawa:2019amx,Haba:2019qol,Dror:2019syi,Blasi:2020wpy,Dunsky:2021tih,Ferrer:2023uwz} focused on the stochastic GW background from local cosmic strings or thermal phase transition dynamics. Here we focus on cosmic strings which associated with {\it global} $U(1)$ symmetry breaking the dynamics of which is essentially very different from that of local cosmic string which lead to novel correlations between GW observables and BSM parameter space quite unexplored before as we will show. Global $U(1)$ breaking can also lead to phase transitions which can under certain circumstances lead to a GW signature \cite{DiBari:2023mwu}, but here we focus on cosmic string signals.
\footnote{It is well known that global symmetries can be broken by gravitational effects, but here we shall assume that gravitational breaking has a negligible effect on the cosmic string dynamics well below the Planck scale.}

%Recently it has been proposed to complement these indirect tests with the observations of GWs of primordial origin 
%and the try to understand the implications of neutrino physics for the first time in this regard. %Moreover depending on whether the broken symmetry is of global or local (gauged) the cosmic string configurations are different and consequently the GW spectrum arising from them
%\ag{Need to add citations in this paragraph.}

$U(1)_{B-L}$ symmetry is one of the most appealing $U(1)$ extensions of the Standard Model (SM). Although baryon number ($B$) and lepton number ($L$) are both accidental global symmetries of the Standard Model, their difference, $B-L$, is the only anomaly-free combination \cite{Wilczek:1979hc,Lipkin:1980rg,Heeck:2014zfa}. Also $B-L$ symmetry is not only preserved by SU(5) gauge interactions but also protected by sphaleron processes. In general, $B-L$ can be either a global symmetry \cite{Marshak:1980dg,Mohapatra:1982aj,Mohapatra:1982xz} or a local (gauged) symmetry \cite{Davidson:1978pm,Mohapatra:1980qe,Wetterich:1981bx,Buchmuller:1991ce}. The gauged $U(1)_{B-L}$ is more popular in model building as it can be a residue symmetry of the SO(10) group in Grand Unified Theories (GUTs) \cite{Fukuyama:2004ps}. The connection between gauged $U(1)_{B-L}$ symmetry breaking and gravitational waves has also been discussed wildly in literature \cite{Dror:2019syi,King:2020hyd,King:2021gmj,Fu:2022lrn,Dasgupta:2022isg,Okada:2020vvb}. However, the gauged $U(1)_{B-L}$ extension of the SM requires the addition of three right-handed neutrinos so that the gauge anomalies are cancelled. In seesaw models with only two RH neutrinos, the $U(1)_{B-L}$ symmetry can only exist as a global symmetry. The connection between {\it global} $U(1)_{B-L}$ symmetry and gravitational waves has not so far been studied in the type I seesaw framework.

In this paper, then, we make a first study of the connection between neutrino physics and gravitational waves sourced from the dynamics of {\it global} cosmic strings. By considering a global $U(1)_{B-L}$ extension of the type I seesaw model, the $U(1)_{B-L}$ symmetry breaking is related to the mass of the heaviest RH neutrino up to an undetermined Yukawa coupling. After the $U(1)_{B-L}$ symmetry is broken by a heavy scalar, the RH neutrinos become massive and in the meantime global cosmic string network are formed. The evolution of the global strings produce stochastic gravitational waves background (SGWB) that can be detected via several upcoming GW experiments. Such consideration provides us a probe to the mass scale of the heaviest RH neutrino in the type I seesaw model. 

As an example of the general approach to probing the type I seesaw at high scales using GW signals, we study a particular global $U(1)_{B-L}$ extension of an existing model in the literature known as the Littlest Seesaw model. By fitting both the low energy neutrino data and the baryon asymmetry of the universe via leptogenesis, we determine the favoured best fit values for both RHN masses appearing in this model. In particular, by updating the data and improving the numerical method in Ref. \cite{King:2018fqh}, we evaluate the goodness of fitting for different values of the {\it heavier} RHN mass, whose fit is dominated by the low energy neutrino data. We remark that the {\it heavier} RHN mass is mainly relevant for the $U(1)_{B-L}$ breaking scale and GWs, while the {\it lighter} RHN mass is mainly relevant for leptogenesis. 
Choosing regions around the best fitted point, we show how the experimental sensitivity reaches from the GW detectors may be used to probe the mass of the heavier RHN mass in this model, which is predicted from low energy neutrino data. Moreover, we also identify the parameter space which is already ruled out due to existing constraints on global cosmic strings (limits on the string tension G$\mu$) coming from the CMB measurements which we describe in detail.

This paper is organised as follows. In Sec.\ref{sec:model}, we describe the Littlest Seesaw model with a global $U(1)_{B-L}$ symmetry. By revisiting the Littlest Seesaw model, we show how the parameters other than the heaviest RH neutrino mass can be fixed by the neutrino data and leptogenesis. In Sec.\ref{sec:cs}, we briefly review the property of gravitational wave produced by the evolution of cosmic string. After that, we show how the gravitational wave can be used to test the neutrino mass models in Sec.\ref{sec:gw}, with an example of the best-fit benchmark point in the Littlest seesaw model.  Finally, we summarise and discuss in Sec.\ref{sec:con}.

%%%%%%%%%%%%%%%%%%%%%%%%%%%%%%%%%%%%%%%%%%%
\section{Type I seesaw model with a $U(1)_{B-L}$ symmetry \label{sec:model}}
%%%%%%%%%%%%%%%%%%%%%%%%%%%%%%%%%%%%%%%%%%%
\begin{table}[t!]
\centering
\begin{tabular}{ccccccccc}
\hline \hline \\[-9pt]
& $Q$ & $u_R$ & $d_R$ & $L$ & $e_R$ & $H$ & $N$  & $\Phi$  \\[1pt] \hline \\[-9pt]
$SU(2)_L$ & $\bf 2$ & $\bf 1$ & $\bf 1$ & $\bf 2$ & $\bf 1$ & $\bf 2$ & $\bf 1$  & $\bf 1$  \\ [3pt]
$U(1)_Y$ & $\frac{1}{6}$ & $\frac{2}{3}$ & $-\frac{1}{3}$ & $-\frac{1}{2}$ & $-1$ & $-\frac{1}{2}$ & 0 & 0 \\ [3pt]  
& \qquad\qquad & \qquad\qquad & \qquad\qquad & \qquad\qquad & \qquad\qquad & \qquad\qquad & \qquad\qquad & \qquad\qquad \\ [-8pt]
$U(1)_{B-L}$ & $\frac13$ & $\frac13$ & $\frac13$ & $-1$ & $-1$ & $0$ & $-1$ & $2$ \\[2pt] \hline \hline
\end{tabular}
\caption{\label{tab:model} \it Irreducible representations of the fields of the model under the $SU(2)_L \times U(1)_Y \times U(1)_{B-L}$ symmetry. The fields $Q, L$ are left-handed SM doublets while $u_R,d_R,e_R$ are right-handed SM singlets. $N$ represents the right-handed neutrinos and $\Phi$ is a scalar singlet.}
\end{table}

Here, we start with a type I seesaw extension of the SM with a $U(1)_{B-L}$ symmetry. The particle content of the model is shown in Tab.\ref{tab:model} In the frame work of type I seesaw model, the SM leptons $L_\alpha$ couple to two or three singlet fermions $N_i$, namely the right-handed neutrinos, and the Higgs boson through Yukawa-like interactions that can be written as
\begin{eqnarray}
Y_{\alpha i} \overline{L_\alpha} \tilde{H} N_i + h.c..
\end{eqnarray}
The right-handed neutrinos are assumed to be Majorana so that the SM left-handed neutrinos can obtain effective Majorana masses at low scale after the electroweak symmetry breaking. The model is free of anomalies even if the $U(1)_{B-L}$ symmetry is gauged in the case with three RH neutrinos, but with the absence of the third RH neutrino, the model only admits a global $U(1)_{B-L}$ symmetry unless the symmetry is flavour-dependent. 

The Majorana mass of right-handed neutrinos can be sourced from the vacuum expectation value (VEV) of a scalar singlet, which couples to the right-handed neutrinos in the form of
\begin{eqnarray}
\frac12 y_i \Phi \overline{N_i^c} N_i + h.c. \label{eq:RHN_mass}
\end{eqnarray}
As the RH neutrinos are charged under the hypothetic $U(1)_{B-L}$ symmetry, the scalar singlet has to be also charged and thus its VEV $\langle\Phi\rangle=\eta$ would break the symmetry spontaneously. After the $U(1)_{B-L}$ symmetry is broken, the RH neutrinos become massive with a diagonal mass matrix
\begin{eqnarray}
\begin{pmatrix} M_1 & 0 & 0\\ 0 & M_2 & 0 \\ 0 & 0 & M_3 \end{pmatrix}, \quad M_i=y_i \eta\,.
\end{eqnarray}
In models only two RH neutrinos, the heavy neutrino mass matrix is 
\begin{eqnarray}
\begin{pmatrix} M_1 & 0 \\ 0 & M_2  \end{pmatrix}, \quad M_i=y_i \eta\,.
\end{eqnarray}

\subsection{The Littlest Seesaw model}

To make a testable connection between the low energy neutrino physics and the high energy gravitational wave phenomena, we consider a class of highly predictive models, which is called the Littlest Seesaw models. In the type I seesaw model with two right-handed (RH) neutrinos, the neutrino Dirac mass is denoted by $3\times2$ matrix $m_D$. Under the assumption of the constrained sequential dominance (CSD) \cite{King:2015dvf}, the two columns of $m_D$ follow specific alignments. The first column of $m_D$ is proportional to $(0,1,1)$, and the second column is proportional to 
$(1, n, n-2)$. Let $m_a$ and $m_b$ be the coefficients of the two columns, then the Dirac mass matrix can be expressed as 
\begin{eqnarray}
m_D = 
\begin{pmatrix} 0 & m_b e^{i\beta/2} \\ m_a & n\, m_b\, e^{i\beta/2} \\ m_a & (n-2)m_b\, e^{i\beta/2} \end{pmatrix}\,. \label{eq:mass_dirac}
\end{eqnarray}
The relative Majorana phase between the two columns of the Dirac mass matrix, $\beta$, is equivalent to that between the two RH neutrino masses. The Dirac neutrino mass matrix is originated from the neutrino Yukawa coupling through Higgs mechanism. The neutrino Yukawa coupling reads
\begin{eqnarray}
Y_\nu = m_D/v_{\rm SM}=
\begin{pmatrix} 0 & b\, e^{i\beta/2} \\ a & n\, b\, e^{i\beta/2} \\ a & (n-2)b\, e^{i\beta/2} \end{pmatrix}\,, \label{eq:Yukawa_dirac}
\end{eqnarray}
where $a=m_a/v_{\rm SM}$, $b=m_b/v_{\rm SM}$ and $v_{\rm SM}$ is the standard model Higgs VEV. 

In the Littlest Seesaw model, the neutrino mass is explained by the type I seesaw mechanism. In general, the SM neutrino mass matrix is completely determined by 6 parameters in the model: $n$, $\beta$, $m_a$, $m_b$, $M_1$, $M_2$. However, by reprameterisation, the number of independent free parameters can be reduced to 4, namely $n$, $\beta$, $M_a\equiv m_a^2/M_1$, $M_b\equiv m_b^2/M_2$. Using the first alignment in Eq.\ref{eq:mass_dirac} as an example, one can obtain the neutrino mass matrix as
\begin{eqnarray}
m_\nu =
M_b\, e^{-i\beta} \begin{pmatrix} 
1 & n & n-2 \\ 
n & n^2 + r^{-1}\,e^{i\beta} & n(n-2)+r^{-1}\,e^{i\beta} \\ 
n-2 & n(n-2)+r^{-1}\,e^{i\beta} & (n-2)^2+r^{-1}\,e^{i\beta}
\end{pmatrix}\,, \label{LHN_mass}
\end{eqnarray} 
where we further define $r= M_b/M_a$ for convenience. Moreover, the value of number $n$ is commonly motivated by discrete flavour symmetries. We first treat $n$ as a free parameter and then consider a specific case with $n=3$ (motivated by a $S_4$ flavour symmetry) in numerical analysis, which has been studied a lot not only in theoretical aspect but also in the context of leptogenesis \cite{King:2018fqh} and dark matter \cite{Chianese:2019epo,Chianese:2020khl}. 
The mass matrix should be diagonalised by the Pontecorvo-Maki-Nakagawa-Sakata (PMNS) mixing matrix, which involves 3 mixing angles and 1 Dirac phase that are measured by neutrino experiments and 1 relative Majorana phase (not 2 as the lightest neutrino is massless), and have eigenvalues $(0,\,\sqrt{\Delta m^2_{12}},\,\sqrt{\Delta m^2_{13}})$. With the oscillation data \cite{Esteban:2020cvm,NuFitweb}, the parameters $\beta$, $M_a$, $M_b$ can be determined. To evaluate the discrepancy between predicted and measured values of the observables, we adopt the $\chi^2$ function as a measurement, with the definition 
\begin{eqnarray} 
\chi^2 = \sum_n \chi^2_n= \sum_n \left[ \frac{{\cal O}_n^{\rm pre} - {\cal O}_n^{\rm bf}}{\sigma_{{\cal O}_n}} \right]^2 \,.
\end{eqnarray} 
For an observable ${\cal O}_n$, ${\cal O}_n^{\rm pre}$ is the value predicted by the model and ${\cal O}_n^{\rm bf}$ is the best-fitted value from the data with $1\sigma$ error $\sigma_{{\cal O}_n}$. The data used for the best-fitted values and $1\sigma$ errors are listed in Table.\ref{tab:data}.
\begin{table}[tbp] 
\begin{center}
\begin{tabular}{ c|c c }
\hline \hline 
& \quad Without SK atmospheric data \quad& \quad With SK atmospheric data \quad \\ \cline{2-3} \\[-6pt]
$\theta_{12}/\degree$ & $33.41^{+0.75}_{-0.72}$ & $33.41^{+0.75}_{-0.72}$\\ [3pt]
$\theta_{23}/\degree$ & $49.1^{+1.0}_{-1.3}$ & $42.2^{+1.1}_{-0.9}$\\[3pt]
$\theta_{13}/\degree$ & $8.54^{+0.11}_{-0.12}$ & $8.58^{+0.11}_{-0.11}$\\[3pt]
$\delta/\degree$ & $197^{+42}_{-25}$ & $232^{+36}_{-26}$\\[3pt]
$\frac{\Delta m^2_{12}}{10^{-5}\text{eV}^2}$ & $7.41^{+0.21}_{-0.20}$ & $7.41^{+0.21}_{-0.20}$\\[3pt]
$\frac{\Delta m^2_{13}}{10^{-3}\text{eV}^2}$ & $2.511^{+0.028}_{-0.027}$ & $2.507^{+0.026}_{-0.027}$ \\[3pt]\hline \\[-6pt]
$Y_{\Delta B}/10^{-11}$ & \multicolumn{2}{c}{$8.72^{+0.04}_{-0.04}$}\\[3pt]
\hline\hline
\end{tabular}
\end{center}
\caption{\it Global fit result to oscillation data provided by NuFit \cite{Esteban:2020cvm,NuFitweb} and baryon asymmetry from combined analysis of the Planck CMB power spectra, CMB lensing reconstruction and baryon acoustic oscillation (BAO) measurements \cite{Planck:2018vyg}.  \label{tab:data}}
\end{table}
As the true data does not follow the normal distribution, the fitting result has different upper and lower error at $1\sigma$ level. To simplify the calculation of $\chi^2$, we set the $1\sigma$ error to be the one with smaller absolute value in the upper and lower error.

As the neutrino Dirac mass and the RH neutrino mass cannot be determined by the neutrino mass and mixing, we take leptogenesis into consideration. The lepton asymmetry produced during thermal leptogenesis is affected by the lightest RH neutrino mass. In the case of hierarchical right-handed neutrino mass spectrum ($M_2\gg M_1$), the leptogenesis can be estimated as
\begin{eqnarray}
Y_{\Delta \alpha} \simeq \epsilon_{1\alpha} \eta_\alpha Y_{N_1}^{\rm eq}
\end{eqnarray}
where $\epsilon_{1\alpha}$ is the CP asymmetry in the decay of the lightest right-handed neutrino into lepton flavour $\alpha$, $\eta_\alpha$ is the efficiency factor, and $Y_{N_1}^{\rm eq}\simeq4 \times 10^{-3} $ is the equilibrium comoving density of the same neutrino at $T\ll M_1$. The CP asymmetry arising at one-loop order reads
\begin{eqnarray}
\epsilon_{1\alpha} &\simeq& \frac{3}{16\pi}\frac{M_1 M_b}{v_{\rm SM}^2}  \sin\beta \left\{0,\, n(n-1) ,\, (n-1)(n-2)\right\}\,.
\end{eqnarray}
The key mass-dimension parameters in determination of $\eta_\alpha$ appears to be independent of the RH neutrino mass:
\begin{eqnarray} 
\tilde{m}_{\alpha\alpha} &\equiv& \frac{m_a^2\left\{0,\, 1 ,\, 1\right\}_{\alpha}\left\{0,\, 1 ,\, 1\right\}_\alpha}{M_1} = M_a \left\{0,\, 1 ,\, 1\right\}_\alpha \,, \\
m^* &\equiv& 4\pi\frac{v_{SM}^2}{M_1^2}\Hub(M_1)\simeq 1.1\times10^{-3}\, \text{eV}\,,\label{eq:m_quantities}
\end{eqnarray}
where $\Hub(M_1)$ is the Hubble parameter when $T=M_1$. By requiring a successful leptogenesis, $M_1$ can be determined. 

\begin{figure}[t!]
\begin{center}
\includegraphics[width=0.7\textwidth]{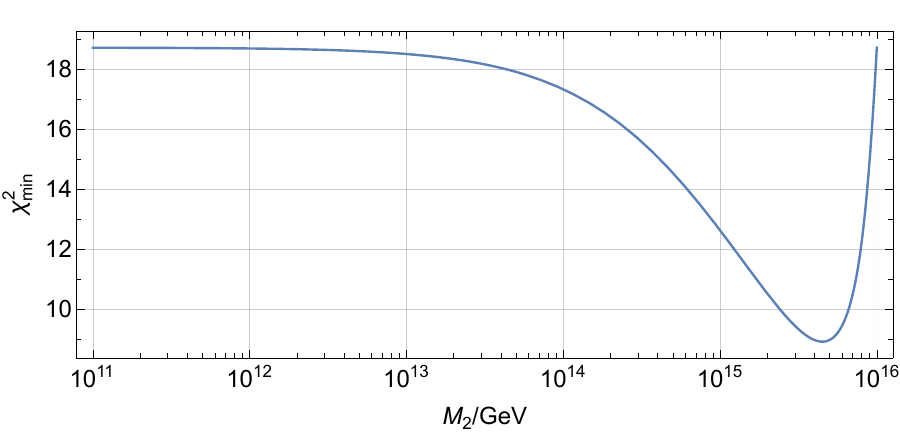}
\caption{\it
Minimal $\chi^2$ for different $M_2$.
We have fitted the results assuming the NuFit data with SK atmospheric data ($\theta_{23}$ in the first octant). The phase $\beta$ is fixed to be $2\pi/3$. The GUT cutoff is set to be $10^{16}$ GeV.
\label{fig:M2Chisq}}
\end{center}
\end{figure}

Although the heaviest RH neutrinos mass $M_2$ does not play an important role in either the low energy neutrino data or the leptogenesis, it can potentially change the behaviour of the renormalisation group (RG) running from low to high scale. As the flavour symmetry is commonly defined at a high scale, the RG running effects should be considered in fitting the model to data. By scanning the parameter space, a value of $M_2$ where the model fit data best can be found in principle. Such a possibility is discussed in \cite{King:2018fqh}. For a benchmark point, it has been shown that local minima exist in the $a-b$ plane as well as the $M_1-M_2$ plane. However, such result does not lead to a local minimum in the 4 parameter space $\{a,\,b,\,M_1,\,M_2\}$. The local minimum in the $a-b$ plane or the $M_1-M_2$ plane only shows that the $2\times2$ blocks of the total Hessian matrix have positive determinant, but the determinant of the total Hessian matrix can still be negative, corresponding to a saddle point. Here, we improve the scan of parameter space using a three-dimensional random walk with random step size for different values of $M_2$. Through the random walk, we fit the model to neutrino data and find the benchmark point that fits the data best. For $\theta_{23}$ in the second octant, we found that there is no local minimum for the $\chi^2$. Instead, the fit becomes worse as $M_2$ increases from $10^{11}$ GeV to $10^{15}$ GeV and then turn back to the same level at GUT scale. However, when $\theta_{23}$ is in the first octant, we find a minimal $\chi^2$ between $10^{11}$ GeV to the GUT scale. In figure Fig.\ref{fig:M2Chisq}, we show how the model fit the data as $M_2$ changes, with respect to the global fit result with SK atmospheric data.
 The model fits the data best when $M_2= 4.50\times10^{15}$ GeV, with $\chi^2=8.93$. The values of the free parameters and the predicted observables for the best-fit point are listed in Tab.\ref{tab:benchmark}.
\begin{table}[t!]
\centering
\begin{tabular}{cccccccccc}
\hline \hline \\[-9pt] 
 $M_1/$GeV & $M_2/$GeV & $a$ & $b$ & $\beta$ & $n$\\[1pt] 
 $5.140\times 10^{10}$ & $4.500\times 10^{15}$ & $0.008617$ & $0.8834$ & $2\pi/3$ & 3 \\ \hline\hline \\[-9pt] 
 $m_2$ & $m_3$ & $\theta_{12}$ & $\theta_{13}$ & $\theta_{23}$ & $\delta$ & $Y_B$ & $\chi^2$ \\[1pt]
$8.70$ meV & $49.98$ meV & $33.89\degree$ & $8.47\degree$ & $44.3\degree$ & $266\degree$ & $8.72\times 10^{-11}$ & $8.93$\\ 
%pulls & $0.44$ & $0.97$ & $5.12$ & $0.64$ & $0.12$ & $1.67$ & $8.64\times10^{-4}$ & \\
\hline \hline 
\end{tabular}
\caption{\label{tab:benchmark} \it Best-fit point found by a three-dimensional random walk with random step size in ${a,\,b,\,M_1}$ for different values of $M_2$. We have fitted the results assuming the NuFit data with SK atmospheric data. }
\end{table}
Among the observables, $\theta_{23}$ shows the greatest deviation from the NuFit result, being outside the $1\sigma$ range. However, the global fit result of $\theta_{23}$ is non-Gaussian. For individual experiments, the combination of ${\theta_{23},\, \delta}$ still lies in the $1\sigma$ range allowed by the T2K result and $2\sigma$ range allowed by the NOvA result \cite{Esteban:2020cvm}.

{\subsubsection{Neutrinoless double beta decay} 

The Majorana nature of neutrinos would lead to neutrinoless double beta decay ($0\nu\beta\beta$).
The key parameter affecting the decay rate (or the half-life time) of nucleons due to the mediation of light Majorana neutrinos is described by the effective mass $m_{\beta\beta}$, which is determined by the mass spectrum and PMNS mixing matrix. 
In the framework of the Littlest Seesaw model, the neutrino mass spectrum and mixing matrix are completely fixed by the neutrino data. 
Following the curve in Fig.\ref{fig:M2Chisq}, the model predicts $m_{\beta\beta}$ around 4 meV, beyond the sensitivity of the next generation experiments \cite{nEXO:2021ujk,LEGEND:2021bnm}, which is consistent with the common result in the normal ordering case with the lightest neutrino massless \cite{Giuliani:2019uno}. 
}

{ \subsubsection{Vacuum stability} 

As the couplings run from low scale to high scale, the Higgs quartic coupling can become negative, leading to breaking of the stability of the vacuum. Within SM, the RG equation of the Higgs quartic coupling at 1-loop level is given by \cite{Degrassi:2012ry} 
%\bof{I left this to you on purpose as you are more familiar with this area, and you may want to cite some of your works.

\begin{eqnarray}
(4\pi)^2\frac{d \lambda_H}{d\ln \mu} &=&
\frac{3}{8} \left(g_1^4 + 2 g_1^2 g_2^2 + 3 g_2^4\right) - 6 y_t^4 + \left(12 y_t^2 - 3 g_1^2 - 9 g_2^2\right)\lambda_H + 
24 \lambda_H^2\,.
\end{eqnarray}
Such running depends significantly on the Yukawa coupling (or the mass) of the top quark since it the largest among other couplings. In the seesaw extension of SM, the seesaw Yukawa couplings contributes negatively to the RG equation \cite{Antusch:2005gp}
\begin{eqnarray}
(4\pi)^2\frac{d \lambda_H}{d\ln \mu} &=&
\frac{3}{8} \left(g_1^4 + 2 g_1^2 g_2^2 + 3 g_2^4\right) - 6 y_t^4 - 2 \text{Tr}[Y_\nu^\dagger Y_\nu Y_\nu^\dagger Y_\nu]  \nonumber \\
&& \quad\quad+ \left(12 y_t^2 + \text{Tr}[Y_\nu^\dagger Y_\nu] - 3 g_1^2 - 9 g_2^2\right)\lambda_H + 
24 \lambda_H^2\,.
\label{eq:RG_NR}
\end{eqnarray}
For heavy RH neutrinos, the seesaw Yukawa coupling can be quite large (close to O(1)), leading to a sharp decrease to values below zero of the SM Higgs quartic coupling at scales above the mass of the heaviest RH neutrino \cite{Mandal:2019ndp}. This makes the SM Higgs potential unbounded from below and makes the vacuum unstable. However, the existence of the new heavy scalar, which as similar mass to the heavy neutrinos, can also couple to the Higgs through $\lambda_{H\Phi} H^2 \Phi^2$, providing an extra contribution to the RG equation.
At 1-loop level, the RG equations are given by \cite{Elias-Miro:2012eoi}
\begin{eqnarray}
(4\pi)^2\frac{d \lambda_H}{d\ln \mu} &=&
\frac{3}{8} \left(g_1^4 + 2 g_1^2 g_2^2 + 3 g_2^4\right) - 6 y_t^4 - 2 \text{Tr}[Y_\nu^\dagger Y_\nu Y_\nu^\dagger Y_\nu]  \nonumber \\
&& \quad\quad+ \left(12 y_t^2 + \text{Tr}[Y_\nu^\dagger Y_\nu] - 3 g_1^2 - 9 g_2^2\right)\lambda_H + 
24 \lambda_H^2 + 4 \lambda_{H\Phi}^2
%4 \lambda_H(3\text{Tr}[Y_d^\dagger Y_d] + 3\text{Tr}[Y_u^\dagger Y_u] + \text{Tr}[Y_e^\dagger Y_e] + \text{Tr}[Y_\nu^\dagger Y_\nu])-
%2 (3\text{Tr}[Y_d^\dagger Y_d Y_d^\dagger Y_d] + 3\text{Tr}[Y_u^\dagger Y_u Y_u^\dagger Y_u] + \text{Tr}[Y_e^\dagger Y_e Y_e^\dagger Y_e] + \text{Tr}[Y_\nu^\dagger Y_\nu Y_\nu^\dagger Y_\nu])
\nonumber\\
(4\pi)^2\frac{d \lambda_{H\Phi}}{d\ln \mu} &=& \frac12 \left(g_1^4 + 2 g_1^2 g_2^2 + 3 g_2^4\right) + 8 \lambda_\Phi^2 + 4 (3\lambda_H + 2 \lambda_\Phi) \lambda_{H\Phi}  \\
(4\pi)^2\frac{d \lambda_\Phi}{d\ln \mu} &=& - \frac18 \sum_i y_i^4 + 8 \lambda_{H\Phi}^2 + 
\sum_i y_i^2 \lambda_\Phi + 20 \lambda_\Phi^2 \nonumber
\end{eqnarray}
The negative contribution from the seesaw Yukawa can be compensated with the positive contribution from the coupling between SM Higgs and the heavy scalar $\lambda_{H\Phi}$, and one may avoid the vacuum instability \cite{Mandal:2019ndp}.  
}

%%%%%%%%%%%%%%%%%%%%%%%%%%%%%%%%%%%%%%%%%%%%%%%%%%%%%%%%%%%%%%%%%%%%%%%%%%
%%%%%%%%%%%%%%%%%%%%%%%%%%%%%%%%%%%%%%%%%%%%%%%%%%%%%%%%%%%%%%%%%%%%%%%%%%

\section{Gravitational Waves from Global Cosmic Strings \label{sec:cs}}

\begin{figure}[t!]
\begin{center}
\includegraphics[width=0.7\textwidth]{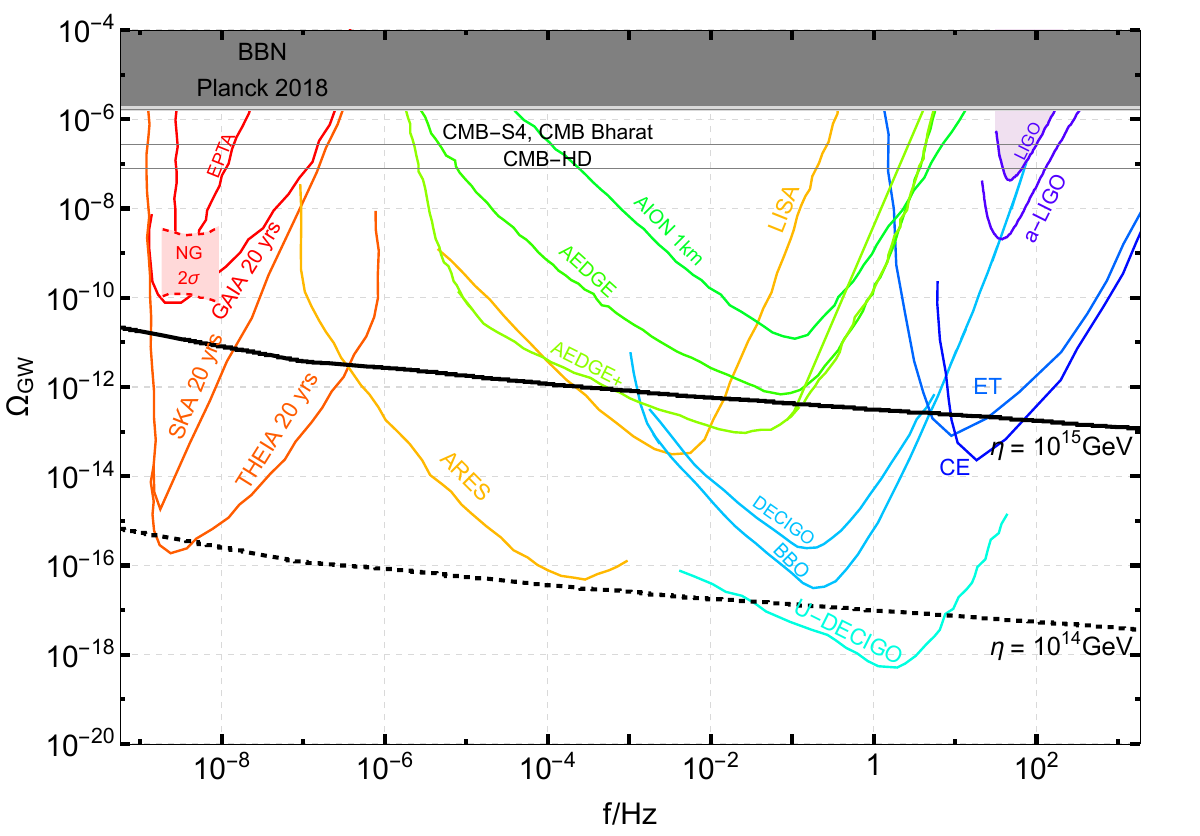}
\caption{\it Gravitational wave spectrum from a global cosmic string network with $U(1)_{\rm B-L}$ symmetry breaking scale $\eta = 10^{14}, 10^{15}$ GeV.}  
\label{fig:GWspectrum}
\end{center}
\end{figure}

Cosmic strings (CS) are topological defects that are produced due to $U(1)$ symmetry breaking in the early universe \cite{Hindmarsh:1994re, Vilenkin:2000jqa, Vachaspati:2015cma}. These topological defects
behave as dynamical classical objects moving at relativistic speed. In context to string theory however, the description of these objects are sometimes as fundamental or somtimes as composite objects \cite{Witten:1985fp, Dvali:2003zj,Copeland:2003bj, Polchinski:2004ia, Sakellariadou:2008ie, Davis:2008dj, Sakellariadou:2009ev, Copeland:2009ga}. Interestingly, CS networks once formed offer very promising sources of GW of cosmological origin which maybe detected in near future. Moreover several Standard Model extensions, such as models of Grand Unified Theory (GUT) \cite{Jeannerot:2003qv,Sakellariadou:2007bv,Buchmuller:2013lra}, or the seesaw mechanism for generating the neutrino masses in the Standard Model when $U_{B-L}(1)$ is broken spontaneously \cite{Dror:2019syi}.

The cosmic string network is characterized by its correlation length $L$. When the strings are stretched by cosmic expansion, they form loops. This would mean we may expect that $L$ evolves linearly with the scale factor $a$ due to the background Hubble expansion, in a manner that  $L\propto t^{1/2}$ during radiation domination epoch and $L \propto t^{2/3}$ during the matter domination. But this turns out to be incorrect. From we find from the simulations of cosmic strings is that the system reaches its scaling regime after a transient evolution. During this period the energy loss of long strings into loop formation, is exactly the same as that as if $L$ scales linearly with the Hubble horizon $t$ \cite{Ringeval:2005kr, Vanchurin:2005pa, Martins:2005es, Olum:2006ix, BlancoPillado:2011dq}. Therefore, the CS evolutionary dynamics during this regime is only characterized by the string tension $\mu$, which is approximately equal to the phase transition temperature $T_{p}$ squared given as
\begin{equation}
G\mu \simeq 10^{-15} \left( \frac{T_{p}}{10^{11}~\text{GeV}} \right)^2.
\end{equation}
The long string energy density, $\rho_{\infty} = \mu/L^2$ redshifts as radiation in radiation domination and as matter in matter domination epochs in early universe in the scaling regime.

The oscillations of the CS loops are known to be large dominant source of the Stochastic Gravitational Waves Background (SGWB). These long-standing source starts to emit GW after the network formation and still radiating today \cite{Vilenkin:1981bx, Hogan:1984is, Vachaspati:1984gt, Accetta:1988bg, Bennett:1990ry, Caldwell:1991jj,Allen:1991bk, Battye:1997ji, DePies:2007bm, Siemens:2006yp, Olmez:2010bi, Regimbau:2011bm, Sanidas:2012ee, Sanidas:2012tf, Binetruy:2012ze, Kuroyanagi:2012wm, Kuroyanagi:2012jf}. The prediction of the SGWB from CS is that these GW consists of frequencies spanning over many orders of magnitude in frequency. Hence, the capability of the next generation of GW interferometers, LISA \cite{Audley:2017drz}, Einstein Telescope \cite{Hild:2010id, Punturo:2010zz}, Cosmic Explorer \cite{Evans:2016mbw}, BBO and DECIGO \cite{Yagi:2011wg} to detect the SGWB from CS. This naturally gives us a unique observational window, on any new physics beyond the SM in early universe like we allude to, in the littlest seesaw model. For our analysis, we will refer to Ref.~\cite{Kibble:1976sj} for the original article, Ref.~\cite{Vilenkin:2000jqa} for a textbook, and Refs.~\cite{Gouttenoire:2019kij,Auclair:2019wcv, Gouttenoire:2022gwi,Ghoshal:2023sfa} for reviews of their spectrum arising due to the GW emission.

\subsection{Short review for Global Cosmic Strings}

The CS core has the size which is inverse of the scale of symmetry-breaking, typically much smaller than the cosmological horizon. Due to this it can be described as infinitely thin classical objects with energy per unit length also knwon as the $\mu$ (Nambu-Goto approximation) for the global cosmic strings,
\begin{equation}
\mu = 2\pi n \, \eta^2 \times 
%1 ~ ~ ~ & {\rm for ~ local ~ strings},\\
\log(\eta t)
\label{string_tension}
\end{equation} 
where $\eta$ represents the vacuum expectation value of the scalar field constituting CS, and $n$ is the winding number (taken to be $n=1$). Usually for global strings, the logarithmic divergence arises due to the presence of massless Goldstone mode. This means the existence of long-range gradient energy \cite{Vilenkin:2000jqa}. As the CS network is formed below the temperature of the $U(1)$-breaking phase transition, the cosmic string tension is approximately given by
\begin{align}
T_\textrm{form} &\simeq \eta. %\simeq 10^{11} \textrm{ GeV}\left(\frac{G\mu}{10^{-15}}\right)^{1/2}.
\label{string_formation_cutoff}
\end{align}
There exists no GW before the cosmic strings network is formed. The typical GW spectrum from global cosmic strings has a natural cut-off, which corresponds to the network formation giving us the frequency,
\begin{align}
f_\textrm{form} ~ \simeq {1 \rm GHz} \times 
0.47 \left(\frac{T_{\rm form}}{10^{14} ~ {\rm GeV}}\right) \left(\frac{0.1}{\alpha}\right) \left[\frac{g_*(T_{\rm form})}{g_*(T_0)}\right]^{\frac{1}{4}},
\label{eq:freq_string_formation_cutoff}
\end{align}
where it has always assumed that the early universe is always radiation-dominated universe. This cut-off remains only in the ultra-high frequency regime, and is not probe-able by the future-planned GW interferometer-based GW detectors.

%\subsection{ GW emission from loops.}

In general, the evolution of cosmic strings is initially frozen due to the presence of thermal friction; however afterwards it reaches an attractor solution called the \emph{scaling regime}. It is during this period that the correlation length of the string network suffers linear growth with respect to the cosmic time, $L \propto t$ \cite{Vilenkin:2000jqa, Martins:2000cs, Martins:2016ois}. Dedicated numerical simulations of cosmic strings \cite{Blanco-Pillado:2013qja} show that the GW spectrum is dominantly produced via loops with the largest size, typically corresponding to $10\%$ of the Hubble horizon size. Although the loop may have some disctribution but the fact that even the largest cosmic string loop size is so very small compared to the Hubble means that we may take the loop size distribution to be monochromatic in nature for all practical purposes
\begin{align}
P_\alpha(\alpha) = \delta(\alpha - 0.1).
\end{align}
After their formation, loops oscillate and radiate GW at a frequency $\tilde{f} = 2k/l$, where $l$ is the loop length and $k \in \mathbb{Z}^+$ denotes the usual Fourier mode index under consideration. The frequency of the GW observed today is given by $f = \tilde{f}[a(\tilde{t})/a(t_0)].$ where the subscript 0 represents present time.
Each Fourier mode $k$ radiates GW with power
\begin{align}
P_{\rm GW}^{(k)} = \Gamma^{(k)} G\mu^2, \qquad \text{with} \quad \Gamma^{(m)} = \frac{\Gamma k^{-{\delta}}}{\sum_{p=1}^\infty p^{-\delta}},
\label{eq:power_emisssion_GW_strings}
\end{align}
where $\Gamma = 50$ \cite{Blanco-Pillado:2017oxo}. The index $\delta$ depends on whether high Fourier modes are dominated by cusps ($\delta = 4/3$), kinks ($\delta =5/3$), and kink-kink collision $(\delta=2)$ \cite{Olmez:2010bi}. 
The strings will lose energy incessantly leading to shrinking of the loop length $l$
\begin{align}
l(\tilde{t}) =  \alpha t_i - (\Gamma G \mu + \kappa) (\tilde{t} - t_i). \label{eq:length_shrink}
\end{align}
where $\Gamma G \mu$ and $\kappa$ represent the shrinking rates due to GW and particle emissions, respectively.
%Local-string loops dominantly decay via the GW emission ($\kappa = 0$), while global-string loops dominantly decay into Goldstone modes with $\kappa = \Gamma_{\rm Gold}/2 \pi \log(\eta t) \gg \Gamma G\mu$ where $\Gamma_{\rm Gold} \approx 65 $ \cite{Vilenkin:1986ku}.
%\end{itemize}

Written from right to left, are in chronological order the various processes that occur and lead us to the final expression for the spectral energy density of GWs from CS, defined as $\Omega_{\rm GW}\equiv \frac{1}{\rho_c}\frac{d\rho_{\rm GW}}{d\ln{f}}$,
\begin{align}
\Omega_{\rm GW}(f) &= \sum_m \frac{1}{\rho_c} \int_{t_{\rm osc}}^{t_0} d\tilde{t} \int_0^1 d\alpha \,\Theta\left[t_i - \frac{l_*}{\alpha}\right]\cdot \Theta[t_i - t_{\rm osc}]\cdot\left[\frac{a(\tilde{t})}{a(t_0)} \right]^4 \cdot P_{\rm GW}^{(m)} \times \nonumber\\
&\hspace{15em} \times\left[\frac{a(t_i)}{a(\tilde{t})} \right]^3 \cdot P_\alpha(\alpha)\cdot \frac{dt_i}{df}\cdot \frac{d n_{\rm loop}}{d t_i}. \label{eq:master_eq_pedagogical}
\end{align}
which simplifies to \cite{Gouttenoire:2019kij} \begin{align}
\Omega_{\rm std}^{\rm CS} h^2 &\simeq \Omega_r h^2 \, \Delta_{\rm R}\,C_{\rm eff}^{\rm rad} \cdot
9 \left[\frac{\Gamma}{\Gamma_{\rm gold}}\right] \left[\frac{\eta}{M_{\rm Pl}}\right]^{4} \log^3\left[ \eta \tilde{t}_{\rm M}\right], 
\\[0.5em]
&\simeq 
1.2 \cdot 10^{-17} \left[\frac{\eta}{10^{15} ~ {\rm GeV}}\right]^{4} \log^3\left\{ 5.6 \cdot 10^{30} \left[\frac{\eta}{10^{15} ~ {\rm GeV}}\right] \left[ \frac{1 ~ {\rm mHz}}{f} \right]^{\! 2} \right\}, 
\label{cs_flat_amp_local}
\end{align}
where $\Omega_{r}h^2 \simeq 4.2 \times 10^{-5}$ \cite{ParticleDataGroup:2020ssz} is the radiation density today, and $\Delta_{\rm R}$ gives the information on the change in Universe expansion rate due to the change of the number of relativistic species,
\begin{align}
\label{eq:Delta_R}
\Delta_{\rm R}(T) \equiv \left( \frac{g_*(T)}{g_*(T_0)}\right)\left(\frac{g_*s(T_0)}{g_{*s}(T)} \right)^{4/3} = ~0.39 \left(\frac{106.75}{g_*(T)} \right)^{1/3}.
\end{align}

\medskip

\subsection{Existing constraints on Global strings\label{l1}}

Massless Goldstone particles can be produced efficiently by global cosmic strings, contributing to the number $N_{\rm eff}$ of effective relativistic degrees of freedom. The precise constraint depends on how many Goldstone particles can be produced from strings, which is still debatable. Very recent studies \cite{Hindmarsh:2019csc,Hindmarsh:2021vih,Buschmann:2019icd,Buschmann:2021sdq} claim that the Goldstone energy spectrum from strings is scale-invariant, while other also recent studies \cite{Gorghetto:2018myk,Gorghetto:2020qws,Gorghetto:2021fsn} suggest a slightly infrared-dominated spectrum, which leads to the production of more Goldstone particles. Here we quote the upper bound $\eta \lesssim 3.5 \cdot 10^{15} \, {\rm GeV}$ derived in Ref.~\cite{Chang:2021afa} and refer to Refs.~\cite{Gorghetto:2021fsn,Dror:2021nyr} for slightly tighter bounds.
%Since this is the leading energy-loss channel, the $\Delta N_{\rm eff}$ constraint from Goldstone bosons is more severe than the one from GW and put a bound on the string scale. 

The measurements of CMB show no evidence of B-mode polarization. This gives us yet another constraint on the global cosmic string network. If one assumes instantaneous reheating and the presence of only the SM degrees of freedom, the upper bound on the primordial inflationary Hubble parameter $H_{\rm inf}\lesssim 3\times 10^{13}~\rm GeV$ \cite{BICEP:2021xfz} roughly gives the estimate for the maximum possible temperature of the universe to be $T_{\rm max} \lesssim 4 \times 10^{15}$~GeV. Therefore, for the string network to form, the string scale $\eta$ must be at least smaller than the maximum temperature $\eta \lesssim 4 \times 10^{15}$~GeV, up to $\mathcal{O}(1)$ model-dependent parameters.

Additional constraints arise due to the potential distortion of the CMB power spectrum by global cosmic strings \cite{Planck:2015fie,Charnock:2016nzm,Lopez-Eiguren:2017dmc}.
For $\eta \gtrsim 10^{15} \, {\rm GeV}$, GW from global strings extend to $f \lesssim 10^{-14} \, {\rm Hz}$. This in principle leaves its signature in CMB polarization experiments, e.g. Ref.~\cite{BICEP:2021xfz}.  However, the GW in this frequency range can only be produced after photon decoupling, evading the CMB constraint. We refer the reader to see Fig.~8 of Ref.~\cite{Chang:2021afa}.

\subsection{GW Detectors}
\noindent 
%We  display the (expected) sensitivity curves for a variety of existing and proposed experiments that can be grouped in terms of 
In Fig.\ref{fig:GWspectrum}, we display the expected sensitivity reaches for various current and planned GW experiments which can be categorized as following:
\begin{itemize}
    \item \textbf{ground based  interferometers:} \textsc{LIGO}/\textsc{VIRGO}             \cite{LIGOScientific:2016aoc,LIGOScientific:2016sjg,LIGOScientific:2017bnn,LIGOScientific:2017vox,LIGOScientific:2017ycc,LIGOScientific:2017vwq}, a\textsc{LIGO}/a\textsc{VIRGO}  
    \cite{LIGOScientific:2014pky,VIRGO:2014yos,LIGOScientific:2019lzm}, \textsc{AION} \cite{Badurina:2021rgt,Graham:2016plp,Graham:2017pmn,Badurina:2019hst}, \textsc{Einstein Telescope (ET)} \cite{Punturo:2010zz,Hild:2010id}, \textsc{Cosmic Explorer (CE)}  \cite{LIGOScientific:2016wof,Reitze:2019iox},
    \item   \textbf{space based interferometers:}  \textsc{LISA} \cite{LISA:2017pwj,Baker:2019nia}, \textsc{BBO} \cite{Crowder:2005nr,Corbin:2005ny}, 
    \textsc{DECIGO}, \textsc{U-DECIGO}\cite{Seto:2001qf,Kudoh:2005as,Nakayama:2009ce,Yagi:2011wg,Kawamura:2020pcg}, \textsc{AEDGE} \cite{AEDGE:2019nxb,Badurina:2021rgt}, \textsc{$\mu$-ARES} \cite{Sesana:2019vho}
    \item \textbf{CMB spectral distortions:} \textsc{PIXIE}, \textsc{Super-PIXIE}  \cite{Kogut:2019vqh}, \textsc{VOYAGER2050}~\cite{Chluba:2019kpb}
    \item \textbf{recasts of star surveys:} \textsc{GAIA}/\textsc{THEIA} \cite{Garcia-Bellido:2021zgu}, 
    \item \textbf{CMB polarization:} Planck 2018 \cite{Akrami:2018odb} and BICEP 2/ Keck \cite{BICEP2:2018kqh} computed by  \cite{Clarke:2020bil}, \textsc{LiteBIRD} \cite{Hazumi:2019lys}, 
     \item \textbf{pulsar timing arrays (PTA):} Square-Kilometer-Array (\textsc{SKA}) \cite{Carilli:2004nx,Janssen:2014dka,Weltman:2018zrl}, \textsc{EPTA} \cite{Lentati:2015qwp,Babak:2015lua}, \textsc{NANOGRAV}~\cite{McLaughlin:2013ira,NANOGRAV:2018hou,Aggarwal:2018mgp,Brazier:2019mmu,NANOGrav:2020bcs}
\end{itemize}

\subsection{Dark radiation bounds from BBN and CMB decoupling}

At last the energy density of the primordial gravitational waves should be smaller than the limit on dark radiation encoded in $\Delta N_\text{eff.}$ from Big Bang Nucleosynthesis and CMB observations (see the discussion in the text for bounds and projections on $\Delta N_\text{eff.}$). The change of the number of effective
relativistic degrees of freedom (Neff ) at recombination
time is given by an amount \cite{Maggiore:1999vm}
\begin{align}
    \int_{f_\text{min}}^{f=\infty} \frac{\text{d}f}{f}   \Omega_\text{GW}(f) h^2 \leq 5.6\times10^{-6}\;\Delta N_\text{eff.}.\label{eq:darkrad}
\end{align}

The lower limit for the integration is taken to be $f_\text{min}\simeq 10^{-10}\text{Hz}$ for BBN and $f_\text{min}\simeq 10^{-18}\text{Hz}$ for the CMB bounds. 
%In practice, when \textit{e.g.} plotting many GW spectra simultaneously, and as a first estimate we may 
However we may approximately ignore the frequency dependence to constrain the energy density of the peak for a given GW spectrum to be
\begin{align}
    \Omega_\text{GW}^\text{Peak} h^2 \leq   5.6\times10^{-6}\;\Delta N_\text{eff.}.\label{eq:darkrad2}
\end{align}

%%%%%%%%%%%%%%%%%%%%%%%%%%%%%%%%%%%%%%%%%%%
\section{Gravitational wave from $U(1)_{B-L}$ symmetry breaking  \label{sec:gw}}
%%%%%%%%%%%%%%%%%%%%%%%%%%%%%%%%%%%%%%%%%%%

As having been discussed in Sec.\ref{sec:model}, the scale of $U(1)_{B-L}$ symmetry breaking can be related to the masses of RH neutrinos through a single group of Yukawa couplings, namely the $y_i$ in Eq.\ref{eq:RHN_mass}. On the other hand, GW detectors can detect the gravitational waves produced by the cosmic strings resulting from the $U(1)_{B-L}$ symmetry breaking, whose strength is dominantly determined by the scale of $U(1)$ symmetry breaking. As a consequence, it is possible to constraint the masses of RH neutrinos through gravitational wave detection, up to the Yukawa coupling $y_i$. In particular, if a certain upper bound of the Higgs singlet VEV is obtained from gravitational wave observation, the heaviest RH neutrino, which has the largest Yukawa coupling to the heavy Higgs singlet, cannot be much heavier than that upper bound due to the perturbativity limit of the coupling. 

\begin{figure}[t!]
\begin{center}
\includegraphics[width=1\textwidth]{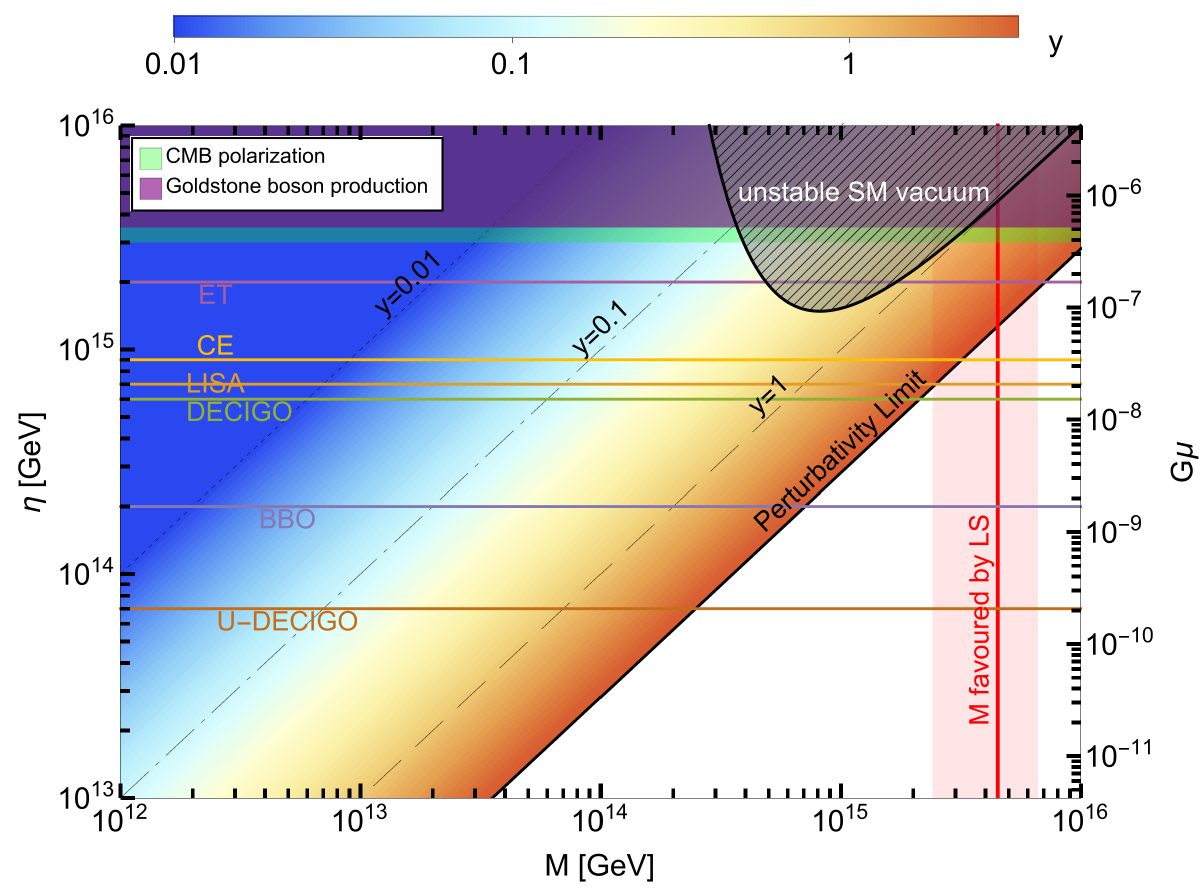}
\caption{\it Constraints and sensitivities in the $M-\eta$ plane. The various GW detectors shown here tell us the range of $U(1)_{\rm B-L}$ symmetry breaking scales that it will be able to probe via the measurements of the GW spectrum.
The horizontal scale represents the heaviest RH neutrino mass, and the lines and contours show the Yukawa coupling of this neutrino to the $B-L$ breaking scalar, whose VEV is indicated by the LH vertical axis.
The pink band is the favoured heavier RH neutrino mass in the Littlest Seesaw model, which can be fully probed by future GW experiments.
See the text for further details. \label{fig:mass_vev}}
\end{center}
\end{figure}

In Fig.\ref{fig:mass_vev}, we show how this kind of connection between RH neutrino mass and the GW observations works. The heaviest RHN mass is labelled as $M$ and the corresponding coupling to the Higgs singlet is $y$. The colour filled in the figure stands for the value of $y$. Below the black solid line, the white region is where the coupling is larger than its perturbativity limit and thus not considered theoretically. The constraints on the VEV of the Higgs singlet are shown are the shadowed areas and the sensitivities of GW detectors are shown as the horizontal lines. In particular, if any of the GW detectors does not find any signal, the region above the corresponding line would be excluded. Combined the perturbativity limit, the GW detection can be used to test models with the heaviest RH neutrino mass above $2.5 \times 10^{14}$ GeV. 

As an example, the case of the Littlest Seesaw model fitting to the NuFit result with SK atmospheric data is presented in Fig.\ref{fig:mass_vev}. The red vertical line marks the value of the heaviest RH neutrino mass in the Littlest Seesaw model for the best-fit benchmark point in Tab.\ref{tab:benchmark}. The region shadowed with red represents the allowed range of the heaviest RH neutrino mass requiring $\chi^2<10$. As can be read from the figure, the region where the Littlest Seesaw model can fit the neutrino data with $\chi^2<10$ can be excluded if no GW signal is observed by ET, CE and LISA. Further more, if all of the GW detectors in the figure cannot find and signal, the Littlest Seesaw model would be excluded at $\chi^2<16.0$ level. If the coupling $y$ is further required to be smaller than 1, the exclusion would be improved to $\chi^2<17.6$. 

On the other hand, the GW detection in Fig.\ref{fig:mass_vev} can be alternatively understood as constraints on the coupling $y$. 
For the best-fitted value of $M_2$ in the Littlest Seesaw model (the red line), ET can imply the coupling $y$ if it is below 5. 
In another word, if ET does not find any signal, the region where $y<5$ will be ruled out for the best-fitted point in the Littlest Seesaw model. 
If a model predicts a lower mass of the heaviest RH neutrino, then its coupling to the superheavy scalar field can be implied or constrained by more GW detection. 
For example, the coupling $y$ in a model predicting a $7\times 10^{13}$ GeV heaviest RH neutrino would be implied by U-DECIGO as long as it is smaller than 1.

{As the right-handed neutrino mass becomes larger, the Yukawa coupling can be large enough to dominate over the top quark Yukawa coupling in the RG running of the Higgs quartic coupling \ref{eq:RG_NR}. As a result, the SM vacuum can become unstable at high energies. In Fig.\ref{fig:mass_vev}, we identify the region where the SM vacuum stability breaks below the scale of the heavy scalar $\eta$ since the heavy scalar is integrated out of the theory. Above the scale of the heavy scalar, the coupling between $\Phi$ and Higgs boson can help preserving the stability of the SM vacuum.}

%\ag{following which set of Eqns. was this analysis done let us specify.}

\medskip

%%%%%%%%%%%%%%%%%%%%%%%%%%%%%%%%%%%%%%%%%%%
\section{Conclusion\label{sec:con}}
%%%%%%%%%%%%%%%%%%%%%%%%%%%%%%%%%%%%%%%%%%%

The type I seesaw model not only accounts for the small neutrino masses and large mixing of the PMNS matrix elegantly but also provides a potential explanation of the matter-antimatter asymmetry via thermal leptogenesis. However, in the standard leptogenesis scenario, the lightest RH neutrino is typically above $10^9$ GeV, which is far beyond the accessibility of collider or astrophysical experiments. 

In this paper, we have explored the possibility of constraining the RH neutrino mass with primordial gravitational wave detection. In the minimal natural extension of the type I seesaw models with a {\it global} $U(1)_{B-L}$ symmetry, the VEV of a superheavy scalar field, from which the RH neutrinos obtain their Majorana mass, breaks the $U(1)_{B-L}$ symmetry. During the corresponding phase transition in the early universe, global cosmic strings can be produced due to the $U(1)_{B-L}$ symmetry breaking. The dynamical evolution of the strings results in detectable GWs, whose amplitude can be determined by the symmetry breaking scale. As a result, the detection of GWs can be used to constrain the masses of heaviest RH neutrinos associated with the ${B-L}$ breaking. 

In some models the RH neutrino masses are determined, leading to a decisive test of these models using GWs. As a concrete example, we have studied the Littlest Seesaw model, where only two RH neutrinos play a role in the seesaw mechanism and the Yukawa couplings in the flavour basis follow special alignments as required by discrete flavour symmetry. By fitting the model to neutrino data and baryon asymmetry, all of the free parameters in the model can be determined including the heavier RH neutrino mass, which is related to the $B-L $ breaking scale up to an arbitary Yukawa coupling.
We have found that, due to the perturbativity limit of this coupling, the parameter space favoured by the Littlest Seesaw model can be fully probed by the proposed GW detectors including LISA, CE and ET. If no GW signal is found by these detectors, the entire parameter space of this model would be disfavoured. 
For more general type I seesaw models with a global $B-L$ symmetry, the above GW detectors can serve to constrain the coupling between the heaviest RH neutrino and the $B-L$ breaking scalar. 

In summary, gravitational wave detection allows us to probe the heaviest RH neutrino mass in  a general class of type I seesaw models with a global $U(1)_{B-L}$ symmetry. To illustrate this, we have analysed a specific example of a highly predictive type I seesaw model with two RH neutrinos and shown that it will be tested very soon by proposed gravitational wave detectors. The methodology can be extended to other type I seesaw models with a global $U(1)_{B-L}$ symmetry. 

In future, it would be interesting to understand how such $U(1)_{\rm }$ global symmetries when embedded in UV-complete scenarios like SO(10) may lead to associated formation of other topological defects like domain walls and local cosmic strings or hybrid defect scenarios which may have their own unique GW signal corresponding to breaking pattern in the early universe like studied in Ref. \cite{Barman:2022yos,Dunsky:2021tih} or involving mixed GW signals from phase transitions and topological defects in standard and non-standard cosmological histories \cite{Ferrer:2023uwz,Ghoshal:2023sfa}. 

We envisage that the precision measurements that the GW cosmology and GW astronomy aspire to reach from the planned global network of GW detectors will make the dream of testing high-scale physics and fundamental BSM scenarios of UV-completion a reality in the very near future.

\section*{Acknowledgments}
BF acknowledges the Chinese Scholarship Council (CSC) Grant No.\ 201809210011 under agreements [2018]3101 and [2019]536. SFK acknowledges the STFC Consolidated Grant ST/L000296/1 and the European Union's Horizon 2020 Research and Innovation programme under Marie Sklodowska-Curie grant agreement HIDDeN European ITN project (H2020-MSCA-ITN-2019//860881-HIDDeN). SFK also thanks IFIC, University of Valencia, for hospitality.

\appendix
\section{Appendix: Signal-to-noise ratio (SNR)}

 Gravitational Wave Interferometers actually measure displacements of detector arms in terms of a what is known in terms of dimensionless strain-noise $h_\text{GW}(f)$ which is related to the GW amplitude. This can be straight-forwardly converted into the corresponding  energy density \cite{Garcia-Bellido:2021zgu}
\begin{align}
    \Omega_\text{exp}(f) h^2 = \frac{2\pi^2 f^2}{3 H_0^2} h_\text{GW}(f)^2 h^2,
\end{align}
where $H_0 = h\times 100 \;\text{(km/s)}/\text{Mpc}$ is the Hubble expansion rate today. 
The signal-to-noise ratio (SNR) for a projected experimental sensitivity $\Omega_\text{exp}(f)h^2$ is then estimated in order to assess the detection probability of the primordial GW background originating from the global cosmic string background following the prescription~\cite{Thrane:2013oya,Caprini:2015zlo}
\begin{align}
     \text{SNR}\equiv \sqrt{\tau \int_{f_\text{min}}^{f_\text{max}} \text{d}f \left(\frac{ \Omega_\text{GW}(f) h^2}{\Omega_\text{exp}(f) h^2}\right)^2 } \label{eq:SNR},
\end{align}
where $h=0.7$ and  $\tau = 4\; \text{years}$ is the observation time. Usually this is chosen to be $\text{SNR}\geq 10$ as the detection threshold for each individual detector.

%%%%%%%%%%%%%%%%%%%%%%%%%%%%%%%%%%%%%%%%%%%
\bibliographystyle{JHEP}
\bibliography{Ref}
\end{document}